\documentclass[12pt]{article}
\pdfoutput=1
\usepackage{latexsym, graphicx,cite,mathrsfs}
\usepackage{amsmath}
\usepackage{amssymb}
\usepackage{amsthm}
\usepackage{subcaption}

\usepackage[top = 1. in,bottom = 1 in, left = 1 in, right=1 in]{geometry}
\usepackage[colorlinks=true, linkcolor=blue, bookmarks=true]{hyperref}
\newcommand{\arXiv}[1]{\href{http://www.arXiv.org/abs/#1}{arXiv:#1}}
\newcommand{\doi}[2]{\href{http://dx.doi.org/#2}{#1}}

\allowdisplaybreaks

\makeatletter
\renewcommand\section{\@startsection {section}{1}{\z@}%
                  {-3.5ex \@plus -1ex \@minus -.2ex}
                  {2.3ex \@plus.2ex}%
                  {\normalfont\large\bfseries}}
\renewcommand\subsection{\@startsection{subsection}{2}{\z@}%
                   {-3.25ex\@plus -1ex \@minus -.2ex}%
                   {1.5ex \@plus .2ex}%
                   {\normalfont\bfseries}}
\makeatother

\newcommand{\beq}{\begin{equation}}
\newcommand{\eeq}{\end{equation}}
\newcommand{\ber}{\begin{array}}
\newcommand{\eer}{\end{array}}
\newcommand{\del}{\partial}

\newcommand{\bea}{\begin{eqnarray}}
\newcommand{\eea}{\end{eqnarray}}
\newcommand{\nn}{\nonumber}

\newcommand{\SD}{Schwinger-Dyson}

\begin{document}
\begin{titlepage}
\begin{flushright}
\phantom{arXiv:yymm.nnnn}
\end{flushright}
\vspace{5mm}
\begin{center}
{\LARGE\bf Schwinger-Dyson recursion, Gr\"obner\vspace{3mm}\\ elimination and graph combinatorics}\\
\vskip 17mm
{\large Oleg Evnin}
\vskip 7mm
{\em High Energy Physics Research Unit, Department of Physics, Faculty of Science, Chulalongkorn University,
Bangkok, Thailand}
\vskip 3mm
{\em Theoretische Natuurkunde, Vrije Universiteit Brussel (VUB) and\\
International Solvay Institutes, Brussels, Belgium}
\vskip 7mm
{\small\noindent {\tt oleg.evnin@gmail.com}}
\vskip 20mm
\end{center}
\begin{center}
{\bf ABSTRACT}\vspace{3mm}
\end{center}

Schwinger-Dyson equations are a central topic in quantum field theory (QFT). Derived using integration by parts, they capture relations between correlators of different orders, forming an open, infinite-dimensional system. In zero-dimensional QFTs with polynomial actions, which is just a physical interpretation for multivariate Gaussian integrals with polynomial insertions, all higher-order correlators (moments) can be expressed in terms of a fixed `master' set. An effective tool for identifying this master set is provided by conventional Gr\"obner elimination. One consequence is a closed finite-dimensional system of differential equations for the coupling dependence of these `master' correlators.

An important application of zero-dimensional QFTs is in counting discrete geometric structures (as in simplicial approaches to quantum gravity).
For the problem of counting labelled regular graphs, applying this Schwinger-Dyson closure results in a computational breakthrough relative to the existing state-of-the-art:
for 9- and 10-regular graphs on $n$ vertices, a finite-depth, finite-width recursive formulation is obtained, making it possible to compute the total graph counts
in $O(n)$ arithmetic operations.

\vfill

\end{titlepage}

\section{Introduction}

It is due to Feynman's insight that we started representing Gaussian integrals with polynomial insertions as drawings of graphs.
These Feynman diagrams went on to become a ubiquitous language in particle physics, high-energy theory and beyond,
and more broadly, an emblematic symbol of modern theoretical physics as such.
The original context was in {\it quantum field theory} (QFT) defined through functional integrals, but ordinary multivariate Gaussian
integrals evaluated in terms of Wick contractions can be represented in this manner equally well.

Because of Feynman's graphic representation, crosstalk between QFT problems and graph combinatorics emerged early on during the relevant historical developments \cite{Hurst,KNN,Cvitanovic78,BIPZ}.
It is not surprising that the correspondence can be turned around: if the language of graphs can be used as a shorthand
for Gaussian integral computations, one would naturally think that problems in discrete geometry and the associated combinatorics 
such as graph enumeration can be usefully recast in terms of Gaussian integrals \cite{BIZ}. A recognizable application of this idea
is in simplicial quantum gravity and related subjects, where matrix and tensor integrals are used to build and count spatial discretizations
\cite{GM,Ambjoern,kazakov,higherd,SasakuraSato}. Such ideas can also be applied to more conventional combinatorial problems,
as in enumeration of labelled simple regular graphs (graphs on distinguishable vertices without self-loops and multiple edges where all vertices have the same number of edges attached).
We will turn at a later point in this exposition to such a Gaussian-integral representation for the number of $d$-regular graphs on $n$ vertices
developed using statistical physics methods in \cite{EH}, based on the earlier considerations in \cite{Kawamoto}, and concurrently and independently by purely combinatorial methods in \cite{CdP23,dP24}.

The applications to discrete geometry and combinatorics outlined above typically rely on `zero-dimensional' (0d) QFT:
there is no spacetime, no momentum loop integral, and the `partition function' is simply an ordinary finite-dimensional integral over $k$ real (or contour-deformed) variables $\mathbf{x}=(x_1,\dots,x_k)$,
\beq\label{Z0d}
Z(t)=\int d\mathbf{x}\, e^{S(t,\mathbf{x})},
\eeq
where the `action' $S(t,\mathbf{x})$ is a polynomial in both the integration variables $\mathbf{x}$ and the `coupling' parameter $t$. 
For readers without QFT backgrounds, the integral (\ref{Z0d}) is all that needs to be known about our basic setup (without making the optional connections
to the historical QFT context).
We are not worried about the convergence of this integral, but rather about the availability of its expansion as
 a formal power series in $t$,
\beq\label{Ztseries}
Z(t)=\sum_{n=0}^{\infty}\, c_n \,\frac{t^n}{n!},
\eeq
and about our control over the coefficients $c_n$. For example, very often, $S(t,\mathbf{x})$ is of the `Gaussian-anchored' form $-{\mathbf{x}^2}/{2}+tV(\mathbf{x})$. Then, expanding $e^{tV}$ as a Taylor series yields each $c_n$ as a convergent Gaussian integral.

It is natural to expect that recursive structures will emerge when evaluating $c_n$ of (\ref{Ztseries}), as one can see in the following simple example.
Consider (\ref{Z0d}) with $S=-{\mathbf{x}^2}/{2}+tV(\mathbf{x})$, and extend the definition somewhat by introducing the auxiliary 
 `sources' $\mathbf{y} = (y_1,\dots,y_k)$,
\begin{equation}\label{eq:Zty}
Z(t,\mathbf{y}) = (2\pi)^{-k/2} \int d\mathbf{x}\, \exp\!\left(-\frac{\mathbf{x}^2}{2} + t V(\mathbf{x}) + \mathbf{y}\cdot\mathbf{x}\right)
= e^{\mathbf{y}^2/2}\sum_{n=0}^{\infty} p_n(\mathbf{y})\frac{t^n}{n!},
\end{equation}
where $\mathbf{x}^2 = \sum_{i=1}^k x_i^2$, $\mathbf{y}^2 = \sum_{i=1}^k y_i^2$, $\mathbf{y}\cdot\mathbf{x} = \sum_{i=1}^k y_i x_i$, and $V(\mathbf{x})$ is a polynomial of total degree $M$. Factoring out the free Gaussian source dependence $e^{\mathbf{y}^2/2}$ ensures that every $p_n(\mathbf{y})$ is a polynomial in $\mathbf{y}$, with $c_n = p_n(\mathbf{0})$ recovering the $t$-expansion of $Z(t,0)$ which is our original objective. Differentiating \eqref{eq:Zty} with respect to $t$ brings down a factor of $V(\mathbf{x})$ inside the integral, which can be replaced by the differential operator $V(\nabla_{\mathbf{y}})$ acting outside the integral, where $\nabla_{\mathbf{y}} = (\partial/\del{y_1},\dots,\partial/\del{y_k})$:
${\partial Z(t,\mathbf{y})}/{\partial t} = V(\nabla_{\mathbf{y}})\, Z(t,\mathbf{y})$.
Using the identity $\partial\big(e^{\mathbf{y}^2/2} f(\mathbf{y})\big)/\del{y_i} = e^{\mathbf{y}^2/2}(y_i + \partial/\del{y_i})f(\mathbf{y})$ to commute $V(\nabla_{\mathbf{y}})$ past $e^{\mathbf{y}^2/2}$ and matching powers of $t^n/n!$ on both sides yields an immediate recursion for the polynomials $p_n(\mathbf{y})$:
\begin{equation}\label{eq:pnrec}
p_{n+1}(\mathbf{y}) = V(\mathbf{y} + \nabla_{\mathbf{y}})\, p_n(\mathbf{y}), \qquad p_0(\mathbf{y}) =1.
\end{equation}
Such basic recursive relations are common in the QFT literature, also for $\log Z$, the generating function of connected diagrams, and the `effective action' $\Gamma$, which is the Legendre transform of $\log Z$ with respect to $\mathbf{y}$. (A very recent reincarnation of more sophisticated relatives of this simple source-polynomial recursion can be seen in \cite{Schwarz}.)

While this recursion on polynomials is completely explicit, it suffers from a severe practical drawback when one wants the zero-source numbers $c_n = p_n(\mathbf{0})$ at large $n$. Because multiplying by $\mathbf{y}$ raises the polynomial degree by one at each application, one concludes that $\deg p_n = Mn$, while a general $k$-variable polynomial of degree $Mn$ contains $\binom{Mn+k}{k} = O(n^k)$ monomials. Computing $c_n$ via \eqref{eq:pnrec} therefore requires tracking an ever-expanding vector of polynomial coefficients whose width grows as $O(n^k)$, with the recursion costing $O(n^{k+1})$ operations rather than $O(n)$ as one would optimally hope.

This brings us to a central question insofar as practical applications are concerned: \emph{how can one convert a growing-width source recurrence like (\ref{eq:pnrec}) into a fixed-width linear recurrence with polynomial coefficients in $n$?} Equivalently, in terms of the original generating functions, how does one find an ordinary linear differential equation in $t$ with polynomial coefficients satisfied by $Z(t)$? This is known to be possible, by general elimination theory, for polynomial actions $S(t,\mathbf{x})$, but the fact remains rather little-appreciated in the QFT literature, despite some related specialized developments that we will briefly recall below. In the computer algebra literature, the question is tackled by the general machinery of holonomic functions and `creative telescoping' \cite{Zeilberger90,SS19}, with techniques involving noncommutative Gr\"obner bases in Weyl algebras and Ore elimination \cite{Stanley,Lipshitz,CMS05,Kauers,BCLS18,CM26,BCL26}. Those general-purpose algorithms apply to broad classes of multivariate integrals and symmetric-function scalar products. In application to $d$-regular graph enumeration, they have been highly successful yielding finite-depth scalar recursion for $d$  up to 7 in  \cite{CM26} and up to 8 in \cite{BCL26}. Yet, these approaches, with their essential need for handling noncommutative structures, eventually hit a computational wall and have so far not been extended beyond $d=8$.

For 0d polynomial QFTs, however, everything simplifies dramatically thanks to the Schwinger-Dyson equations, which we proceed to quickly review. In the integral (\ref{Z0d}), assuming decay at infinity in a sense appropriate for giving definite values to the formal power series coefficients, the integral of any total derivative vanishes identically:
\beq\label{SDintro}
\int d\mathbf{x}\, \frac{\del}{\del x_i}\Bigl[f(\mathbf{x})\,e^{S(t,\mathbf{x})}\Bigr]=0,
\eeq
for any polynomial $f(\mathbf{x})$.
Expanding the derivative produces a linear relation between integrals of different polynomials: the leading terms come from bringing down $\del S/\del x_i$, which raises the polynomial degree, while the derivative acting on a monomial $f(\mathbf{x})=x_1^{p_1}\cdots x_k^{p_k}$ produces lower-degree terms. 
One thus ends up with relations between the `expectation values' of different monomials, which can be called `moments' or `correlators':
\beq
\langle x_1^{p_1}\cdots x_k^{p_k}\rangle_t=\int d\mathbf{x}\, x_1^{p_1}\cdots x_k^{p_k}\,e^{S(t,\mathbf{x})}.
\eeq
Naively, these relations do not form a closed system, connecting any given monomial to higher-degree monomials, and this is how the matter is typically presented in QFT textbooks.

The essence of this exposition is to apply elimination theory and use the \SD{} equations to express the expectation values of all monomials through a finite set
of `master moments' or `master correlators.' How can this happen? Imagine we can show that, for all monomials higher than a certain total degree threshold, the \SD{} equations can be used to lower the degree. If that is the case, the degree can always be pushed down to the threshold, and there is always a finite number of distinct monomials up to any given degree.

The way all of this is implemented goes through the powerful machinery of Gr\"obner bases \cite{CLO15} that we will briefly review in a pedagogical manner when its turn comes.
One key structural feature is that we will only need ordinary Gr\"obner bases for polynomials of commuting variables, built around the operation of polynomial multiplication alone, despite the fact that differentiation is present among the operations inside the \SD{} equations. The essence (to be clarified in the technical exposition below) is that the \SD{} equations tell us that the expectation value of $f\del S/\del x_i +\del f/\del x_i$ is zero. But the only pieces of degree higher than $f$ are contained in the first term  $f\del S/\del x_i$ made of products of ordinary commutative polynomials, while the second term may be neglected for the purpose of identifying the relations that lower the degree, and incorporated as a minor technicality later. This leads one to the problem of
constructing the ordinary Gr\"obner basis of the polynomials $\del S/\del x_i$ (which a physicist would visualize as the left-hand sides of the `classical equations of motion'
$\del S/\del x_i = 0$).  From this Gr\"obner basis, one will be able to proceed with identification of the `master moments' obtaining a reduction of an arbitrary monomial to this master set, and therefore a closure of the equations for the $t$-derivatives of this finite set of master moments, resulting in a linear first-order ODE system with polynomial coefficients. This, in turn, yields a finite-depth, finite-width recursion for the moments that we are aiming to construct, leading to a mechanism for evaluating the expansion coefficients $c_n$ of the partition function $Z$ using $O(n)$ algebraic operations.

While the exposition here differs strongly in its style and techniques from mainstream QFT, it is useful to point out a few specific contacts with more advanced branches of the QFT literature: in works on Feynman diagram accounting, resurgence and large-order behavior \cite{Cvitanovic78,ZinnJustin,Borinsky17}, nonperturbative Lefschetz-thimble decompositions and transseries \cite{ASV12,DunneUnsal14}, recursions among connected Green's functions \cite{Molinari05,PH07}, master integrals for Feynman diagrams \cite{Laporta,Lee,coactions}, intersection-theory constructions in QFT \cite{MM19,Frellesvig19} and the theory of integral discriminants \cite{MSh}. The elementary framework presented here complements these lines of inquiry: rather than relying on invariant theory for purely homogeneous polynomials, constructing intersection pairings, or solving large systems of integration-by-parts identities as in multi-loop momentum integrals, we turn the linear Schwinger-Dyson identities of a multi-variable polynomial action into an automated commutative Gr\"obner elimination algorithm, yielding exact, fixed-width linear recurrences that generate high-order perturbative series in $O(n)$ operations. It is a matter of hope that this simple, transparent picture will come to be of use in the context of other related pursuits.

There are two parts in this essay: First, we present a self-contained, pedagogical derivation of the Schwinger-Dyson closure algorithm for general $0$d polynomial QFTs, keeping the mathematics elementary and transparent. Second, we apply the method to the exact enumeration of labelled $d$-regular graphs via the Gaussian representation of \cite{EH}. In doing so, we reach beyond the existing state-of-the-art and obtain finite-depth, finite-width recursion for regular graph counting up to $d=10$.


\section{Schwinger-Dyson closure via Gr\"obner elimination}
\label{sec:SD-grobner}

Consider a general 0d polynomial QFT, which is a physicist's name for the following integral over $k$ variables $\mathbf{x} = (x_1,\dots,x_k)$ and a single coupling parameter $t$:
\begin{equation}\label{eq:Z-general}
Z(t) = \int_\Gamma d\mathbf{x}\, e^{S(t,\mathbf{x})},
\end{equation}
where the `action' $S(t,\mathbf{x})$ is a polynomial in $\mathbf{x}$ and $t$, and $\Gamma$ is a $k$-dimensional integration contour (typically $\mathbb{R}^k$, though contour deformations can be used whenever needed to ensure convergence). For any polynomial observable $\mathcal{O}(\mathbf{x})$, we denote its unnormalized expectation value (a \emph{moment} or \emph{correlator}) by
\begin{equation}\label{eq:correlator-def}
\langle \mathcal{O}(\mathbf{x}) \rangle_t \equiv \int_\Gamma d\mathbf{x}\, \mathcal{O}(\mathbf{x})\, e^{S(t,\mathbf{x})},
\end{equation}
so that the `partition function' itself is $Z(t) = \langle 1 \rangle_t$. (Properties of integrals like (\ref{eq:Z-general}) have in fact been explored in depth from the point of view of abstract mathematics, see for instance \cite{Adolphson}, and it will be useful to convert this knowledge into practical computational recipes.)

Our objective is to find a finite list of monomials $\mathbf{b}(\mathbf{x}) = (b_1(\mathbf{x}),\dots,b_{\mathcal{R}}(\mathbf{x}))^T$---which we call the \emph{master monomials} (or \emph{master basis})---such that the vector of {\it master correlators} (or {\it master moments})
\begin{equation}\label{eq:master-vec}
\mathbf{v}(t) = \begin{pmatrix} \langle b_1(\mathbf{x}) \rangle_t \\ \vdots \\ \langle b_{\mathcal{R}}(\mathbf{x}) \rangle_t \end{pmatrix}, \qquad b_1(\mathbf{x}) \equiv 1,
\end{equation}
contains all information about the expectation values of all other monomials. Since the action $S$ is polynomial, a consequence is that $\mathbf{v}$ 
satisfies a closed, first-order linear differential system with coefficients polynomial in $t$:
\begin{equation}\label{eq:gm-ode}
D(t)\,\frac{d\mathbf{v}(t)}{dt} = N(t)\,\mathbf{v}(t),
\end{equation}
where $D(t)$ is a nonzero scalar polynomial in $t$ and $N(t)$ is an $\mathcal{R}\times\mathcal{R}$ matrix of polynomials in $t$. Indeed, differentiating $\langle b_i\rangle$
with respect to $t$ yields  $\langle b_i \,\del S/\del t\rangle$, and for a polynomial $S$, it is just a combination of higher-degree monomials that can be expressed back through the master basis. Once \eqref{eq:gm-ode} is known together with the free-theory initial vector $\mathbf{v}(0)$, expanding $\mathbf{v}(t) = \sum_{n=0}^{\infty} \mathbf{v}_n\, t^n/n!$ converts \eqref{eq:gm-ode} directly into a fixed-width linear recurrence for the coefficient vectors $\mathbf{v}_n$.

How is this reduction to master monomials realized in practice? We start by writing out the Schwinger-Dyson identities: for any polynomial $f(\mathbf{x})$, integrating $\partial\big(f(\mathbf{x})e^{S(t,\mathbf{x})}\big)/\del{x_i}$ over $\Gamma$ gives
\begin{equation}\label{eq:SD-op}
\langle \mathcal{D}_i f(\mathbf{x}) \rangle_t = 0, \qquad
\mathcal{D}_i \equiv \frac{\partial}{\partial x_i} + E_i(t,\mathbf{x}), \qquad
E_i(t,\mathbf{x}) \equiv \frac{\partial S(t,\mathbf{x})}{\partial x_i},
\end{equation}
for $i = 1,\dots,k$. Here $\mathcal{D}_i$ is the $i$-th Schwinger-Dyson operator, while the polynomials $E_1,\dots,E_k$ are the left-hand sides of the classical equations of motion $\partial S/\partial x_i = 0$. As alluded to in the introduction, $\partial/\partial x_i$ always lowers the polynomial degree, so for the purpose of identifying combinations of Schwinger-Dyson equations that lower the degree, we may temporarily ignore the derivative piece and treat $\mathcal{D}_i$ as multiplication by $E_i(\mathbf{x})$. In this simplified picture, the approximate identity $\langle E_i(\mathbf{x}) f(\mathbf{x}) \rangle_t \approx 0$ lets one trade the highest-degree term in $E_i$ for lower-degree terms. To make this reduction systematically reach every sufficiently high-degree monomial, it is natural to look for polynomial combinations of $E_1,\dots,E_k$ whose leading terms are as simple as possible --- ideally pure powers of individual variables.

The existence of such combinations is controlled by the joint roots of the classical equations of motion
\begin{equation}\label{eq:classical-eom}
E_1(\mathbf{x}) = E_2(\mathbf{x}) = \cdots = E_k(\mathbf{x}) = 0
\end{equation}
via Hilbert's Nullstellensatz and its extensions, while their practical construction relies on Gr\"obner bases \cite{CLO15}. Specifically, when $\mathbf{x} = \mathbf{0}$ is the unique complex solution of \eqref{eq:classical-eom}, Hilbert's Nullstellensatz guarantees that some power $x_j^{\mu_j}$ ($\mu_j \ge 1$) can be expanded in terms of $E_j$:
\begin{equation}\label{eq:nullstellensatz-power}
x_j^{\mu_j} = \sum_{l=1}^{k} C_{jl}(\mathbf{x})\, E_l(\mathbf{x}).
\end{equation}
Moreover, if $\mathbf{x} = \mathbf{0}$ remains the unique common zero even when only the highest-degree terms of each $E_l$ are retained (preventing top-degree cancellations between different $E_l$), the cofactors $C_{jl}(\mathbf{x})$ can be chosen with no degree overshoot: $\deg(C_{jl} E_l) \le \mu_j$, and hence $\deg C_{jl} < \mu_j$.
Then, for any monomial divisible by $x_j^{\mu_j}$ --- written as $x_j^{\mu_j} P(\mathbf{x})$ --- we can restore the full Schwinger-Dyson operators \eqref{eq:SD-op} inside $\sum_{l=1}^{k} \langle \mathcal{D}_l\big(C_{jl}(\mathbf{x}) P(\mathbf{x})\big) \rangle_t = 0$ to obtain
\begin{equation}\label{eq:heuristic-ibp}
\langle x_j^{\mu_j} P(\mathbf{x}) \rangle_t
= \sum_{l=1}^{k} \left\langle C_{jl}(\mathbf{x}) E_l(\mathbf{x}) P(\mathbf{x}) \right\rangle_t
= -\sum_{l=1}^{k} \left\langle \frac{\partial}{\partial x_l}\big(C_{jl}(\mathbf{x}) P(\mathbf{x})\big) \right\rangle_t.
\end{equation}
Since $\deg C_{jl} < \mu_j$, every term on the right-hand side of \eqref{eq:heuristic-ibp} has degree at most $\deg C_{jl} + \deg P - 1 < \mu_j + \deg P = \deg(x_j^{\mu_j} P)$, expressing $\langle x_j^{\mu_j} P(\mathbf{x})\rangle_t$ strictly through correlators of lower degree. 

\subsection{Elementary reduction for generic actions}
\label{subsec:generic-box}

A systematic approach to identifying the master basis utilizes Gr\"obner elimination and will be exposed in the remainder of this section.
It is wise to start, however, with the simplest case that would typically apply to generic polynomial actions, where the Gr\"obner basis is not actually needed.
For specific actions, and certainly in our applications to graph combinatorics in the second half of this article, applying this `Gr\"obner-free' approach
may produce a bloated master basis, or fail to converge to a finite basis at all, but it is still useful to keep this `null case' in mind before proceeding with
the more sophisticated Gr\"obner elimination technology.

To demonstrate the point, we turn once again to the commonly seen `Gaussian-anchored' action:
\beq\label{eq:gaussian-anchored}
S(t,\mathbf{x}) = -\frac{\mathbf{x}^2}{2} + t\,V(\mathbf{x}).
\eeq
Whenever the top-degree homogeneous part of $V$, denoted $V_{top}$, is generic in the sense that $\partial V_{top}/\del {x_1} = \cdots = \partial V_{top}/\del{x_k} = 0$ has $\mathbf{x} = \mathbf{0}$ as its unique complex zero, one can apply the following simple and universal, though not necessarily optimal construction: The master monomials are identified as the following `box' basis
\beq\label{eq:box-basis}
\mathcal{B}_{\mathrm{box}} = \big\{ x_1^{p_1}\cdots x_k^{p_k} : 0 \le p_i \le M - 2 \;\text{ for all }\; i = 1,\dots,k \big\},
\eeq
where $M$ is the degree of $V$.
The $k$ pure powers $x_1^{M-1},\dots,x_k^{M-1}$ are the only monomials of total degree $M - 1$ that lie outside $\mathcal{B}_{\mathrm{box}}$, and therefore the relations $E_j(\mathbf{x})=\del{S}/\del x_j$ can be treated as a $k\times k$ linear system with respect to $x_j^{M-1}$ so as to obtain the linear combinations $\hat E_j=\sum_l C_{jl}E_l=x_j^{M-1}+\left(P\in \mathcal{B}_{\mathrm{box}}\right)$, where $C_{jl}$ are constants.  This is exactly in the spirit of (\ref{eq:nullstellensatz-power}), with the polynomials $C$ replaced by constants, and such relations can be used to systematically replace any power greater than or equal to $M-1$ of a given variable $x_l$ by an expression containing only powers up to $M-2$, and the derivatives will work out as in (\ref{eq:heuristic-ibp}) since the $C$'s are of degree 0. Repeated application of this reduction will push the expectation value of any monomial into the master correlator set. An explicit example of this box-basis reduction for a two-variable even quartic interaction is worked out in Appendix~\ref{app:quartic}.

\subsection{Gr\"obner bases for master correlator reduction}
\label{subsec:grobner-basics}

While \eqref{eq:nullstellensatz-power} already shows that high powers of every variable can be reduced, finding a minimal set of reduction rules that captures \emph{all} polynomial combinations of $E_1,\dots,E_k$ is precisely where Gr\"obner bases \cite{CLO15} become very useful. We first choose an ordering of the variables --- say, $x_1 \succ x_2 \succ \cdots \succ x_k$---and extend it to a \emph{monomial ordering} $\prec$ that respects polynomial degree. For example, in graded reverse lexicographic order, monomials $\mathbf{x}^{\mathbf{p}} \equiv x_1^{p_1}\cdots x_k^{p_k}$ are compared first by total degree $\sum_{i=1}^k p_i$, and ties among monomials of the same degree are broken lexicographically using the chosen variable ordering (rewrite the monomial as a word with $p_k$ copies of {\bf A}, followed by $p_{k-1}$ copies of {\bf B}, etc., then order as in a dictionary, with the `bigger' entries coming later in the list). Every nonzero polynomial $f(\mathbf{x})$ then has a unique \emph{leading monomial} $\operatorname{lm}(f)$, namely the largest monomial appearing in $f$ with respect to $\prec$. Given any list of polynomials $G = \{g_1,\dots,g_s\}$, one can perform multivariate polynomial division of a polynomial $f$ by $G$: whenever the current leading term of $f$ is divisible by $\operatorname{lm}(g_j)$ for some $j$, one subtracts the appropriate multiple $c\,\mathbf{x}^{\mathbf{p}} g_j$ to cancel that leading term, repeating until no remaining term is divisible by any $\operatorname{lm}(g_1),\dots,\operatorname{lm}(g_s)$. The leftover polynomial is the \emph{remainder} of $f$ upon division by $G$.

In the raw list $\{E_1,\dots,E_k\}$, two generators can share common variable factors in their leading monomials, so multiplying them by suitable monomials and subtracting cancels their top terms and produces a new polynomial combination $\sum_{i=1}^k h_i(\mathbf{x}) E_i(\mathbf{x})$ whose leading monomial was invisible in $\{\operatorname{lm}(E_1),\dots,\operatorname{lm}(E_k)\}$. Such lowering of the leading monomial degrees is evidently beneficial for the elimination procedure we have in mind. A \emph{Gr\"obner basis} $G = \{g_1,\dots,g_s\}$ for $\{E_1,\dots,E_k\}$ is an enlarged set of polynomial combinations
\begin{equation}\label{eq:gj-cofactor}
g_j(t,\mathbf{x}) = \sum_{i=1}^{k} h_{ji}(t,\mathbf{x})\, E_i(t,\mathbf{x})
\end{equation}
obtained by completing all such cancellations. Explicitly, Buchberger's algorithm builds $G$ iteratively: starting from $G = \{E_1,\dots,E_k\}$, it takes each pair $(g_a,g_b)$ in $G$, multiplies each by the minimal monomial needed to match their leading monomials $\operatorname{lm}(g_a)$ and $\operatorname{lm}(g_b)$, subtracts them to cancel the top terms, and divides the resulting difference by the current list $G$; whenever a nonzero remainder survives, it is appended to $G$ and the process repeats until every pair reduces to zero. Standard computer algebra systems carry out this construction automatically.
The defining property of the output is that \emph{every polynomial combination $\sum_{i=1}^k h_i(\mathbf{x}) E_i(\mathbf{x})$ leaves remainder zero upon division by $G$ (equivalently, its leading monomial is divisible by $\operatorname{lm}(g_j)$ for at least one $g_j \in G$), and any polynomial $f(\mathbf{x})$ has a unique remainder under division by $G$ regardless of the order in which divisions are performed.} 

\subsection{Step-by-step elimination algorithm}
\label{subsec:elimination-steps}

With the necessary ingredients in place, the construction of \eqref{eq:gm-ode} proceeds in four elementary steps:

\paragraph{Step 1: Differentiating with respect to the coupling.}
Differentiating any master correlator $\langle b_a(\mathbf{x}) \rangle_t$ with respect to $t$ under the integral sign brings down a factor of $\partial S/\partial t$:
\begin{equation}\label{eq:dt-master}
\frac{d}{dt}\langle b_a(\mathbf{x}) \rangle_t = \left\langle b_a(\mathbf{x})\,\frac{\partial S(t,\mathbf{x})}{\partial t} \right\rangle_t, \qquad a = 1,\dots,\mathcal{R}.
\end{equation}
Because $S(t,\mathbf{x})$ is a polynomial, the right-hand side is a linear combination (with coefficients polynomial in $t$) of the correlators $\langle\mathbf{x}^{\mathbf{p}}\rangle_t$, most of which have higher degree than the master monomials $b_1,\dots,b_{\mathcal{R}}$. To close the system, we must express each such higher-degree correlator as a linear combination of $\langle b_1 \rangle_t, \dots, \langle b_{\mathcal{R}} \rangle_t$ whose coefficients are rational functions in $t$.
\paragraph{Step 2: Generating the Schwinger-Dyson identities.}
Applying the Schwinger-Dyson relation \eqref{eq:SD-op} to each monomial $f(\mathbf{x}) = \mathbf{x}^{\mathbf{p}} = x_1^{p_1}\cdots x_k^{p_k}$ yields an infinite family of linear identities among correlators:
\begin{equation}\label{eq:SD-monomial}
0 = \langle \mathcal{D}_i(\mathbf{x}^{\mathbf{p}}) \rangle_t
= \left\langle p_i\, x_1^{p_1}\cdots x_i^{p_i-1}\cdots x_k^{p_k} + \mathbf{x}^{\mathbf{p}} E_i(t,\mathbf{x}) \right\rangle_t, \qquad i = 1,\dots,k.
\end{equation}
Each relation \eqref{eq:SD-monomial} splits into two pieces:
\begin{enumerate}
\item The `classical equation-of-motion' term $\mathbf{x}^{\mathbf{p}} E_i(t,\mathbf{x})$, which is pure polynomial multiplication by $E_i$ in the space of polynomials in $x_1,\dots,x_k$ whose coefficients are rational functions of $t$;
\item The `quantum' term $p_i\, x_1^{p_1}\cdots x_i^{p_i-1}\cdots x_k^{p_k}$, which has strictly lower degree than $\mathbf{x}^{\mathbf{p}}$.
\end{enumerate}
\paragraph{Step 3: Identifying the master basis.}
Fix a variable ordering and graded monomial ordering on polynomials of $\mathbf{x}$ and compute a Gr\"obner basis $G = \{g_1,\dots,g_s\}$ for the polynomials $E_1,\dots,E_k$ at a generic value of $t$. A monomial $\mathbf{x}^{\mathbf{p}}$ is called \emph{irreducible} if it is not divisible by any of the leading monomials $\operatorname{lm}(g_1),\dots,\operatorname{lm}(g_s)$. Let $\mathcal{B} = \{b_1,\dots,b_{\mathcal{R}}\}$ be the set of all irreducible monomials, ordered by increasing degree so that $b_1 = 1$. Whenever $\mathcal{B}$ is finite, it serves as our master basis, and its cardinality $\mathcal{R} = |\mathcal{B}|$ is the rank of the linear differential system \eqref{eq:gm-ode}.
\paragraph{Step 4: Top-down reduction onto the master basis.}
Generalizing the heuristic reduction \eqref{eq:heuristic-ibp}, take any monomial $\mathbf{x}^{\mathbf{q}} \notin \mathcal{B}$. By definition of $\mathcal{B}$, $\mathbf{x}^{\mathbf{q}}$ is divisible by $\operatorname{lm}(g_j)$ for at least one $g_j \in G$, so we can write $\mathbf{x}^{\mathbf{q}} = \mathbf{x}^{\mathbf{p}}\operatorname{lm}(g_j)$ for some exponent vector $\mathbf{p}$. Multiplying \eqref{eq:gj-cofactor} by $\mathbf{x}^{\mathbf{p}}$ and applying the Schwinger-Dyson identity \eqref{eq:SD-monomial} to every monomial in $\mathbf{x}^{\mathbf{p}} h_{ji}(t,\mathbf{x})$ replaces $\langle \mathbf{x}^{\mathbf{p}} g_j \rangle_t$ by a sum of correlators of strictly lower-order monomials:
\begin{equation}\label{eq:topdown-step}
\langle \mathbf{x}^{\mathbf{q}} \rangle_t = \langle \mathbf{x}^{\mathbf{p}}\operatorname{lm}(g_j) \rangle_t = \left\langle \mathbf{x}^{\mathbf{p}}\big(\operatorname{lm}(g_j) - g_j\big) - \sum_{i=1}^{k} \frac{\partial}{\partial x_i}\big(\mathbf{x}^{\mathbf{p}} h_{ji}(t,\mathbf{x})\big) \right\rangle_t.
\end{equation}
Both pieces on the right-hand side of \eqref{eq:topdown-step} are of lower order than $\mathbf{x}^{\mathbf{q}}$: every monomial in $\mathbf{x}^{\mathbf{p}}(\operatorname{lm}(g_j) - g_j)$ is strictly smaller than $\mathbf{x}^{\mathbf{q}}$ in the ordering $\prec$ because we subtracted the leading term of $g_j$, while $\partial_i(\mathbf{x}^{\mathbf{p}} h_{ji})$ has degree $\deg \mathbf{x}^{\mathbf{p}} + \deg h_{ji} - 1$, which is strictly smaller than $\deg \mathbf{x}^{\mathbf{q}} = \deg \mathbf{x}^{\mathbf{p}} + \deg \operatorname{lm}(g_j)$ whenever $\deg h_{ji} \le \deg \operatorname{lm}(g_j)$. The last inequality in fact holds whenever no top-degree cancellations occur among the $E_i$, since each $\deg E_i \ge 1$ and $\deg \operatorname{lm}(g_j) = \max_i(\deg h_{ji} + \deg E_i)$.
 In practice, one never needs to construct the cofactors $h_{ji}$ explicitly: one simply generates all Schwinger-Dyson identities \eqref{eq:SD-monomial} for monomials $\mathbf{x}^{\mathbf{p}}$ up to a cutoff degree $W_{\max}$, treats every monomial $\mathbf{x}^{\mathbf{q}} \notin \mathcal{B}$ as an unknown and every master monomial $b_a \in \mathcal{B}$ as a parameter, and solves the resulting sparse linear system in terms of rational functions of $t$ by Gaussian elimination. Substituting the reduced expressions for $\langle b_a \partial_t S \rangle_t$ back into \eqref{eq:dt-master} and clearing the common denominator $D(t)$ yields the exact matrix differential equation \eqref{eq:gm-ode}.

\subsection{When is closure guaranteed?}
\label{subsec:closure-conditions}

Why does this reduction terminate on a finite master basis $\mathcal{B}$? Returning to our earlier heuristic \eqref{eq:nullstellensatz-power}, the simplest sufficient condition is perfectly elementary: suppose that $\mathbf{x} = \mathbf{0}$ is the unique complex solution to the classical equations of motion $E_1(\mathbf{x}) = \cdots = E_k(\mathbf{x}) = 0$ at generic $t$, and that their highest-degree homogeneous parts also have $\mathbf{x} = \mathbf{0}$ as their only common complex zero.
To see why this guarantees closure, note first that when $\mathbf{x} = \mathbf{0}$ is the unique common zero of $E_1,\dots,E_k$, Hilbert's Nullstellensatz \cite{CLO15} guarantees the expansion \eqref{eq:nullstellensatz-power}: for each variable $x_j$ ($1 \le j \le k$), some pure power $x_j^{\mu_j}$ can be written as $\sum_{l=1}^k C_{jl}(\mathbf{x}) E_l(\mathbf{x})$. By the defining property of a Gr\"obner basis, dividing $x_j^{\mu_j}$ by $G = \{g_1,\dots,g_s\}$ must leave remainder zero, so at the very first step of division at least one $\operatorname{lm}(g_a)$ must divide $x_j^{\mu_j}$, forcing $\operatorname{lm}(g_a) = x_j^{n_j}$ for some $1 \le n_j \le \mu_j$. Consequently, every exponent $p_j$ in an irreducible monomial $x_1^{p_1}\cdots x_k^{p_k} \in \mathcal{B}$ is bounded by $0 \le p_j < n_j$, leaving finitely many master monomials. Moreover, because the highest-degree homogeneous parts of $E_1,\dots,E_k$ share no nonzero common roots, no polynomial combination of them can cancel its top-degree terms to produce a lower-degree leading monomial; thus every cofactor $h_{ji}$ in \eqref{eq:gj-cofactor} satisfies $\deg(h_{ji} E_i) \le \deg g_j$, which ensures that the derivative term $\partial_i(\mathbf{x}^{\mathbf{p}} h_{ji})$ in \eqref{eq:topdown-step} always has strictly lower degree than $\mathbf{x}^{\mathbf{q}} = \mathbf{x}^{\mathbf{p}}\operatorname{lm}(g_j)$.

Two simple extensions cover almost all polynomial actions encountered in practice \cite{CLO15}. First, if $E_1 = \cdots = E_k = 0$ has several isolated solutions rather than $\mathbf{x} = \mathbf{0}$ alone, the same argument still works: instead of a pure power $x_j^{\mu_j}$, a univariate polynomial $P_j(x_j) = x_j^{\mu_j} + \text{(lower powers of } x_j\text{)}$ can be formed from the $E_l$, and its leading term $x_j^{\mu_j}$ still forces a pure power $x_j^{n_j}$ into $\{\operatorname{lm}(g_a)\}$. Second, it often happens that counting total polynomial degree naively fails because different variables naturally carry different weights. For instance, in $x_1^4 + x_2^2$, assigning weight $1$ to $x_1$ and weight $2$ to $x_2$ makes both terms have weight $4$. Measuring degree by such a weighted sum $\sum_j w_j p_j$ restores the no-cancellation property in the weighted sense. We will use such weighted degrees in our practical applications in section~\ref{sec:SD-regular}. Finally, whenever $E_1 = \cdots = E_k = 0$ still has only finitely many complex solutions (so that $\mathcal{B}$ remains finite) but top-degree cancellations occur, an intermediate reduction step may temporarily climb to a higher degree before dropping back down; in such cases, generating Schwinger-Dyson identities up to a moderately higher cutoff degree $W_{\max}$ captures the missing relations automatically. By contrast, if $E_1 = \cdots = E_k = 0$ admits a continuous family of complex solutions (such as a continuous vacuum manifold), the master basis is infinite and the direct reduction does not close unless the continuous critical manifold can be eliminated by applying a symmetry reduction to the underlying variables (as, for example, for the Mexican hat potential).


\section{Gaussian integrals for regular graph enumeration}\label{sec:graphs}
We now turn to our main combinatorial application: counting simple labelled $d$-regular graphs on $n$ vertices (graphs in which every vertex has degree $d$, with no self-loops and no multiple edges). Denote this count by $r_d(n)$, with $r_d(0)=1$ by convention, and let
\beq\label{Rd_egf}
R_d(t)=\sum_{n=0}^{\infty}\, r_d(n)\,\frac{t^n}{n!}
\eeq
be its exponential generating function. The series cannot be resummed for $d\ge 2$ due to the fast growth of $r_d(n)$, but this will not worry us as we are only using it as a bookkeeping device for $r_d(n)$. Since the sum of vertex degrees $dn$ must be even and a simple $d$-regular graph requires at least $d+1$ vertices, $r_d(n)=0$ whenever $dn$ is odd or $1\le n\le d$.

\subsection{The Gaussian integral representation and cycle-index polynomials}
In \cite{EH}, an exact $d$-dimensional Gaussian integral representation was derived for $r_d(n)$. Starting from the contour-integral formula
\beq
r_d(n) = \oint \prod_{v=1}^{n}\frac{dz_v}{2\pi i \,z_v^{d+1}} \prod_{1\le u < v \le n} (1 + z_u z_v),
\eeq
expanding $\log(1+z_u z_v) = \sum_{m=1}^{d} (-1)^{m-1} (z_u z_v)^m / m$ (which is exact upon truncation at $m=d$ because no power higher than $z_v^d$ contributes to the residue), separating the diagonal $u=v$ terms, and applying a Hubbard--Stratonovich transformation with $d$ auxiliary variables $\alpha_1,\dots,\alpha_d$ yields
\beq\label{EH_rdn}
r_d(n) = \frac{1}{\sqrt{(2\pi)^d d!}} \int_{-\infty}^{\infty} d\alpha_1\cdots d\alpha_d \, \exp\left(-\sum_{m=1}^{d}\frac{\alpha_m^2}{2m}\right) \bigl[\mathcal{P}_d (\alpha_1,\dots,\alpha_d)\bigr]^n,
\eeq
where the `interaction' polynomial $\mathcal{P}_d $ is defined by the single-variable residue
\beq\label{Pd_EH}
\mathcal{P}_d (\alpha_1,\dots,\alpha_d) = \frac{1}{2\pi i}\oint \frac{dz}{z^{d+1}} \exp\left(\sum_{m=1}^{d}\frac{i^{m+1}\alpha_m z^m}{m} + \sum_{m=1}^{\lfloor d/2\rfloor}\frac{(-1)^m z^{2m}}{2m}\right).
\eeq
Summing (\ref{EH_rdn}) over $n\ge 0$ with $t^n/n!$ formally gives the 0d QFT partition function
\beq\label{EH_Zt}
R_d(t) = \frac{1}{\sqrt{(2\pi)^d d!}} \int_{-\infty}^{\infty} d\alpha_1\cdots d\alpha_d\, \exp\left(-\sum_{m=1}^{d}\frac{\alpha_m^2}{2m} + t\mathcal{P}_d (\alpha_1,\dots,\alpha_d)\right),
\eeq
understood as a formal power series in $t$.

Expanding the exponential in \eqref{Pd_EH} shows that $\mathcal{P}_d$ is straightforwardly expressed through the {\it cycle-index polynomial} $P_d({x}_1, \dots, {x}_d)$ of the symmetric group $S_d$ \cite{Comtet},
\begin{equation}\label{eq:Pd-cycle}
\mathcal{P}_d(\alpha_1,\dots,\alpha_d) = P_d(\hat{x}_1,\dots,\hat{x}_d),
\end{equation}
evaluated at shifted arguments $\hat{x}_m = i^{m+1}\alpha_m$ for odd $m$ and $\hat{x}_m = i^{m+1}\alpha_m + (-1)^{m/2}$ for even $m$. The standard family $P_k(x_1,\dots,x_k)$ is generated recursively using the identity
\begin{equation}\label{eq:newton-rec}
k\, P_k(x_1, \dots, x_k) = \sum_{m=1}^{k} x_m\, P_{k-m}(x_1, \dots, x_{k-m}), \qquad P_0 = 1.
\end{equation}

It is worthwhile to comment on the relation of the derivation in \cite{EH} to other graph counting formulas. 
In the symmetric-function formalism \cite{CMS05,CM26,Gessel}, letting $\mathrm{p}_m = \sum_{v} z_v^m$ denote the $m$th power-sum symmetric function, $R_d(t)$ is written as an infinite-dimensional Hall inner product
\begin{equation}\label{eq:gessel-inner}
R_d(t) = \left\langle \exp\!\left[\sum_{m \ge 1} \frac{(-1)^{m-1}}{2m}\big(\mathrm{p}_m^2 - \mathrm{p}_{2m}\big)\right],\, e^{t h_d(\mathrm{p}_1, \dots, \mathrm{p}_d)} \right\rangle,
\end{equation}
where $h_d = P_d(\mathrm{p}_1, \dots, \mathrm{p}_d)$ is the complete homogeneous symmetric polynomial of degree $d$. Identifying the power-sum generators $\mathrm{p}_m=\sum_v z_v^m$ with independent Gaussian random variables maps this abstract inner product onto the explicit $d$-dimensional Gaussian integral \eqref{EH_Zt}. An advantage is that, whereas symmetric-function creative telescoping \cite{CMS05,CM26} must manipulate noncommuting operators made of $2d$ generators $(\mathrm{p}_1, \dots, \mathrm{p}_d, \partial_{\mathrm{p}_1}, \dots, \partial_{\mathrm{p}_d})$, the Gaussian integral representation \eqref{EH_Zt} allows us to integrate out half of the variables analytically before any Gr\"obner basis is ever computed, and then rely on a purely commutative Gr\"obner construction.

It is also good to keep in mind the combinatorial considerations of \cite{CdP23,dP24}, where an identical Gaussian representation was derived concurrently with and independently from \cite{EH}. The methods are completely different from the statistical physics techniques of \cite{EH} and maintain a thorough combinatorial interpretation of the derivations at all intermediate steps.

\subsection{Shifting variables to restore weight-homogeneity}
\label{subsec:tadpole-shift}

It is convenient to redefine the variables so as to make the interaction polynomial strictly weight-homogeneous of total weight $d$ in the sense that, for all of its monomials $\prod_{m=1}^{d} x_m^{p_m}$, one has $\sum_{m=1}^{d} m\,p_m = d$. This is achieved by introducing $x_m$ as
\begin{equation}\label{eq:var-shift}
x_m = (-1)^{\lfloor (m+1)/2 \rfloor}\alpha_m - s_m, \qquad
s_m = \begin{cases} i(-1)^{m/2} & \text{if } m \text{ is even}, \\ 0 & \text{if } m \text{ is odd}, \end{cases}
\end{equation}
so that the shifted arguments $\hat{x}_m$ in \eqref{eq:Pd-cycle} simplify to
\begin{equation}\label{eq:xhat-def}
\hat{x}_m = i^{(m-1)\bmod 2} x_m = \begin{cases} x_m & \text{if } m \text{ is odd}, \\ i\,x_m & \text{if } m \text{ is even}. \end{cases}
\end{equation}
This change of variables shifts the constants out of \eqref{Pd_EH} and into the Gaussian measure. The exponent inside \eqref{Pd_EH} becomes $\sum_{m=1}^{d} \hat{x}_m z^m / m$, so one ends up with the pure, unshifted cycle-index polynomial of homogeneous weight $d$:
\begin{equation}\label{eq:Pd-pure}
P_d({x}_1,\dots,{x}_d) = \sum_{p_1 + 2p_2 + \dots + d p_d = d} \frac{1}{p_1!\cdots p_d!} \prod_{m=1}^{d} \left(\frac{{x}_m}{m}\right)^{p_m}.
\end{equation}

The generating function \eqref{EH_Zt} is then
\begin{equation}\label{eq:Rd-shifted}
R_d(t) = \frac{1}{\sqrt{(2\pi)^dd!}}\int d\mathbf{x}\, \exp\!\left( -\sum_{m=1}^{d} \frac{(x_m + s_m)^2}{2m} + t\, P_d(\hat{x}_1,\dots,\hat{x}_d) \right),
\end{equation}
where the shifted Gaussian measure has exact free-theory ($t=0$) mean $\langle x_m \rangle_0 = -s_m$ and variance $\langle (x_m + s_m)^2 \rangle_0 = m$.

Note that while \eqref{eq:Rd-shifted} keeps every Gaussian factor manifestly convergent on $\mathbb{R}^d$, the factors of $i$ in $\hat{x}_{2j} = i x_{2j}$ and $s_{2j} = i(-1)^j$ are purely formal in perturbation theory: rescaling the even-index variables via $x_{2j} \mapsto -i x_{2j}$ maps $P_d(\hat{x}_1,\dots,\hat{x}_d)$ and $s_{2j}$ to real rational quantities while replacing $-(x_{2j} + s_{2j})^2/(4j)$ by $+(x_{2j} + (-1)^j)^2/(4j)$.
In our symbolic and modular computer algebra implementation \cite{repo}, we will use this modified parameterization so that every polynomial coefficient, initial moment, and differential system entry is a rational number without any imaginary parts.


\section{Schwinger-Dyson closure for regular graphs}
\label{sec:SD-regular}

We now apply the Schwinger-Dyson elimination program of Section~\ref{sec:SD-grobner} to the weight-homoge\-neous regular-graph counting function \eqref{eq:Rd-shifted}.

\subsection{Weighted degrees and Gaussian elimination of high-weight variables}
\label{subsec:heavy-light}

Recalling the weight assignment $w(x_m) = m$ ($1 \le m \le d$) from Section~\ref{subsec:tadpole-shift}, any monomial $\mathbf{x}^{\mathbf{p}} = x_1^{p_1}\cdots x_d^{p_d}$ has total weight
\begin{equation}\label{eq:mon-weight}
w(\mathbf{x}^{\mathbf{p}}) = \sum_{m=1}^{d} m\, p_m.
\end{equation}
Because $P_d(\hat{x}_1,\dots,\hat{x}_d)$ is strictly weight-homogeneous of total weight $d$, every variable $x_m$ with $m > k \equiv \lfloor d/2 \rfloor$ can appear in $P_d(\hat{x}_1,\dots,\hat{x}_d)$ at most linearly (since $2m > d$). Consequently, using \eqref{eq:xhat-def}, we can split $P_d(\hat{x}_1,\dots,\hat{x}_d)$ cleanly into a part independent of $x_{k+1},\dots,x_d$ and terms linear in each $x_m$ ($m > k$):
\begin{equation}\label{eq:Pd-split}
P_d(\hat{x}_1,\dots,\hat{x}_d) = P_d^{(0)}(\hat{x}_1,\dots,\hat{x}_k) + \sum_{m=k+1}^{d} L_m(\hat{x}_1,\dots,\hat{x}_{d-m})\, x_m,
\end{equation}
with
\begin{equation}\label{eq:Lm-def}
L_m(\hat{x}_1,\dots,\hat{x}_{d-m}) = \frac{i^{(m-1)\bmod 2}}{m}\, P_{d-m}(\hat{x}_1,\dots,\hat{x}_{d-m})
\end{equation}
depending only on variables of weight at most $d - m \le k$.

Because the high-weight variables $x_{k+1},\dots,x_d$ appear quadratically in the Gaussian measure and linearly in $P_d(\hat{x}_1,\dots,\hat{x}_d)$, we can integrate them out by completing the square:
\begin{equation}\label{eq:gaussian-elim}
\int_{-\infty}^{\infty} dx_m\, \exp\!\left( -\frac{(x_m + s_m)^2}{2m} + t\, L_m x_m \right)
= \sqrt{2\pi m}\,\exp\!\left( -t\, s_m L_m + \frac{m}{2}\, t^2 L_m^2 \right).
\end{equation}
Performing this integration for $m = k+1,\dots,d$ reduces \eqref{eq:Rd-shifted} to a $k$-dimensional integral,
\begin{equation}\label{eq:Rd-eff}
R_d(t) =  (2\pi)^{-k/2} (k!)^{-1/2} \int dx_1\cdots dx_k\, e^{S_{\mathrm{eff}}(x_1,\dots,x_k;t)},
\end{equation}
with the `effective action'
\begin{equation}\label{eq:Seff}
S_{\mathrm{eff}}(\mathbf{x};t) = -\sum_{m=1}^{k} \frac{(x_m + s_m)^2}{2m}
+ t\!\left( P_d^{(0)}(\hat{x}_1,\dots,\hat{x}_k) - \sum_{m=k+1}^{d} s_m L_m \right)
+ \frac{t^2}{2} \sum_{m=k+1}^{d} m\, L_m^2.
\end{equation}

\subsection{Weighted Gr\"obner basis and closure on master correlators}
\label{subsec:weighted-grobner}

On the reduced $k$-variable integral \eqref{eq:Rd-eff}, the Schwinger-Dyson identities \eqref{eq:SD-monomial} for monomials $\mathbf{x}^{\mathbf{p}} = x_1^{p_1}\cdots x_k^{p_k}$ take the explicit form
\begin{equation}\label{eq:SD-reg}
0 = \left\langle \frac{\partial}{\partial x_m}(\mathbf{x}^{\mathbf{p}}) + E_m(t,\mathbf{x})\,\mathbf{x}^{\mathbf{p}} \right\rangle_t, \qquad
E_m(t,\mathbf{x}) \equiv \frac{\partial S_{\mathrm{eff}}}{\partial x_m} = -\frac{x_m + s_m}{m} + \frac{\partial S_{\mathrm{int}}}{\partial x_m},
\end{equation}
for $m = 1,\dots,k$, where $S_{\mathrm{int}}$ denotes the $t$-dependent `interaction' part of \eqref{eq:Seff}. To respect the natural grading of the variables, we fix the variable ordering $x_1 \succ x_2 \succ \cdots \succ x_k$ and equip the polynomials with rational coefficients $\mathbb{Q}[x_1,\dots,x_k]$ with the \emph{weighted graded reverse lexicographic} (\texttt{wdegrevlex}) monomial ordering with weights $(1,2,\dots,k)$. It would be desirable to understand systematically the `classical equations of motion' $E_1 = \cdots = E_k = 0$ that control the truncation of the set of master monomials, but what we know for sure is that our practical implementation terminates
on a finite set for all cases of interest $d=4,\ldots,10$. For even $d$, every term in $S_{\mathrm{eff}}$ has even weight, so odd-weight correlators vanish identically and $\mathcal{B}$ restricts to even-weight irreducible monomials; for odd $d$, both parities occur because $s_{2j} \neq 0$ induces odd-weight linear terms $-s_{2j} L_{2j}$.

Applying the top-down Schwinger-Dyson elimination of Section~\ref{sec:SD-grobner} produces the exact $\mathcal{R}_d \times \mathcal{R}_d$ first-order differential system
\begin{equation}\label{eq:gm-reg}
D_d(t)\,\frac{d\mathbf{v}(t)}{dt} = N_d(t)\,\mathbf{v}(t),
\end{equation}
where $\mathbf{v}(t) = (\langle b_1 \rangle_t,\dots,\langle b_{\mathcal{R}_d} \rangle_t)^T$ has $v_1(t) = \langle 1 \rangle_t = R_d(t)$. In Appendix~\ref{app:d4}, we carry out this entire derivation explicitly for $d = 4$ ($k = 2$), obtaining a rank-$2$ system and the exact recurrence for $r_4(n)$.

\subsection{Vector vs.\ scalar recursion and $O(n)$ evaluation}
\label{subsec:vector-rec}

In the combinatorial literature \cite{CM26,BCL26}, one typically converts the first-order matrix differential system \eqref{eq:gm-reg} into a single scalar linear ODE of order $\mathcal{R}_d$ for $R_d(t)$ (for example, by cyclic-vector elimination) and then translates that ODE into a scalar linear recurrence for $r_d(n)$. While a scalar recurrence is attractively compact, forming it requires  $O(\mathcal{R}_d)$ differentiations of $N_d(t)/ D_d(t)$, which inflates polynomial degrees in $t$ by a factor of $O(\mathcal{R}_d)$ and produces huge integer coefficients.

By contrast, working directly with the first-order vector differential system \eqref{eq:gm-reg} is more economical. Writing $\mathbf{v}(t) = \sum_{n=0}^{\infty} \mathbf{v}_n\, t^n/n!$ --- so that the first component of $\mathbf{v}_n$ is the exact graph count $(\mathbf{v}_n)_1 = r_d(n)$ --- and extracting the coefficients of $t^m/m!$ on both sides of \eqref{eq:gm-reg} yields the direct fixed-width vector recurrence as a system of linear equations:
\begin{equation}\label{eq:vec-rec}
\sum_{j=s}^{\deg D_d} d_j\, m^{\underline{j}}\, \mathbf{v}_{m-j+1} = \sum_{j=0}^{\deg N_d} m^{\underline{j}}\, N_j\, \mathbf{v}_{m-j},
\end{equation}
where $D_d(t) = \sum_{j=s}^{\deg D_d} d_j t^j$ starts with the power $t^s$, $N_d(t) = \sum_{j=0}^{\deg N_d} N_j t^j$ has matrix coefficients $N_j \in \mathbb{Q}^{\mathcal{R}_d \times \mathcal{R}_d}$, and $m^{\underline{j}} = m(m-1)\cdots(m-j+1)$ denotes the falling factorial. Converting this system to scalar recurrence for $r_d(n)$ remains an attractive but optional step, which we will not pursue here, since already at the level of (\ref{eq:vec-rec}), evaluating $r_d(n)$ requires $O(n)$ arithmetic operations for any $d$.

\subsection{Initialization at $t=0$ and indicial boundary conditions}
\label{subsec:init-indicial}

To launch the recurrence \eqref{eq:vec-rec}, we need the free-theory expectation vector $\mathbf{v}_0 = \mathbf{v}(0)$ at $t = 0$. At $t = 0$, the variables $x_1,\dots,x_k$ decouple completely into independent shifted Gaussians with mean $\langle x_m \rangle_0 = -s_m$ and variance $\langle (x_m + s_m)^2 \rangle_0 = m$. Thus, the one-dimensional moments $\mu_m(a) \equiv \langle x_m^a \rangle_0$ satisfy the elementary three-term recurrence
\begin{equation}\label{eq:free-moments}
\mu_m(0) = 1, \qquad \mu_m(1) = -s_m, \qquad \mu_m(a+1) = -s_m\,\mu_m(a) + m\,a\,\mu_m(a-1),
\end{equation}
and every entry of $\mathbf{v}_0$ is simply the product $(\mathbf{v}_0)_a = \prod_{m=1}^{k} \mu_m(p_{a,m})$ for the master monomial $b_a = \prod_{m=1}^{k} x_m^{p_{a,m}}$.

One small subtlety arises when the leading polynomial $D_d(t)$ vanishes at $t=0$ to order $s > 1$ (which occurs for all $d \ge 4$). At $m = s, s+1, \dots$, the coefficient $d_s m^{\underline{s}}$ in \eqref{eq:vec-rec} is nonzero, cleanly determining $\mathbf{v}_{m-s+1}$ from earlier vectors; however, for $1 \le m < s$, the left-hand side of \eqref{eq:vec-rec} vanishes identically, creating an $s$-dimensional indicial nullspace in the lowest Taylor coefficients $\mathbf{v}_1,\dots,\mathbf{v}_{s-1}$. Combinatorially, this nullspace is pinned uniquely without any extra effort: because no simple $d$-regular graph exists on $1 \le n \le d$ vertices, we have $r_d(n) = (\mathbf{v}_n)_1 = 0$ identically for all $1 \le n \le d$. Appending these $d$ exact linear conditions $(\mathbf{v}_n)_1 = 0$ ($1 \le n \le d$) to the initial block of \eqref{eq:vec-rec} determines $\mathbf{v}_1,\dots,\mathbf{v}_d$ uniquely and propagates the exact integer sequence $r_d(n)$ for all $n > d$. In our implementation \cite{repo}, we determine $\mathbf{v}_0,\dots,\mathbf{v}_s$ directly by evaluating the first $s+1$ perturbative coefficients of the $k$-dimensional Gaussian integral \eqref{eq:Rd-eff}, which pins all $t=0$ boundary data uniformly for every $d$ before launching the $O(n)$ recurrence \eqref{eq:vec-rec}.

\subsection{Modular arithmetic and rational reconstruction}
\label{subsec:results}

When $d \ge 6$, performing symbolic Gaussian elimination directly over the rational functions with integer coefficients $\mathbb{Q}(t)$ encounters severe intermediate coefficient swell: rational-function entries inside the reduction matrix develop high-degree numerators and denominators with multi-hundred-digit integer coefficients, even though the final polynomials $D_d(t)$ and $N_d(t)$ are far more compact.

To bypass intermediate expression swell completely, we replace exact $\mathbb{Q}(t)$ elimination with a three-stage modular evaluation--interpolation pipeline (explained in more detail in Appendix~\ref{app:modular}):
\begin{enumerate}
\item \textbf{Finite-field specialization.} Instead of manipulating polynomials in $t$ with rational coefficients, we pick several large $61$-bit prime numbers $q_1,\dots,q_L$ and evaluate the Schwinger-Dyson linear system at $K$ distinct integer points $t = t_1,\dots,t_K \pmod{q_{\ell}}$. Over a finite field $\mathbb{F}_{q_{\ell}}$, every matrix entry is a single machine word, so sparse Gaussian elimination runs at native hardware speed in parallel across CPU cores with zero coefficient growth.
\item \textbf{Chinese Remainder Theorem and rational reconstruction.} For each evaluation point $t_i$, we combine the residues modulo $q_1,\dots,q_L$ via the Chinese Remainder Theorem into an integer modulo $M = \prod_{\ell=1}^{L} q_{\ell}$, and lift that integer to the exact rational entry $M_{ab}(t_i) \in \mathbb{Q}$ of $D_d(t_i)^{-1} N_d(t_i)$ via Euclidean rational reconstruction.
\item \textbf{Univariate rational interpolation.} From the exact rational values $\{M_{ab}(t_i)\}_{i=1}^{K}$, polynomial rational reconstruction recovers each entry of $D_d(t)^{-1} N_d(t) \in \mathbb{Q}(t)$ uniquely, and clearing the least common multiple of denominators yields $D_d(t)$ and $N_d(t)$ over $\mathbb{Z}[t]$. 
\end{enumerate}

Once $D_d(t)$ and $N_d(t)$ are known over $\mathbb{Z}[t]$, unrolling the vector recurrence \eqref{eq:vec-rec} directly over $\mathbb{Q}$ would still require bulky multiprecision fraction arithmetic, both because $r_d(n)$ grows to hundreds of digits and because advancing each step requires dividing by $d_s m^{\underline{s}}$ (and solving a small coupled linear system for the lowest coefficients $\mathbf{v}_1,\dots,\mathbf{v}_d$ where $D_d(t)$ vanishes at $t=0$). Instead, an analogous multi-prime solver evaluates \eqref{eq:vec-rec} modulo $q_1,\dots,q_L$ in parallel using native $64$-bit arithmetic and reconstructs the exact multiprecision integers $r_d(n)$ via the Chinese Remainder Theorem.

\subsection{Results}\label{sec:results}

\begin{table}[t]
\centering
\small
\begin{tabular}{cccccc}
\hline
$d$ & $k = \lfloor d/2 \rfloor$ & Rank $\mathcal{R}_d$ & $\deg D_d$ & $\deg N_d$ & $r_d(n)$ for $n = d+1, d+2, \dots$ \\
\hline
$4$  & $2$ & $2$  & $6$   & $8$   & $1,\; 15,\; 465,\; 19355$ \\
$5$  & $2$ & $6$  & $6$   & $13$  & $1,\; 0,\; 3507,\; 0,\; 66462606$ \\
$6$  & $3$ & $6$  & $15$  & $20$  & $1,\; 105,\; 30016,\; 11180820$ \\
$7$  & $3$ & $20$ & $10$  & $23$  & $1,\; 0,\; 286884,\; 0,\; 480413921130$ \\
$8$  & $4$ & $19$ & $49$  & $58$  & $1,\; 945,\; 3026655,\; 11555272575$ \\
$9$  & $4$ & $70$ & $15$  & $36$  & $1,\; 0,\; 34944085,\; 0,\; 6551246596501035$ \\
$10$ & $5$ & $66$ & $177$ & $191$ & $1,\; 10395,\; 438263364,\; 19506631814670$ \\
\hline
\end{tabular}
\caption{Parameters of the first-order matrix differential system $D_d(t)\mathbf{v}'(t) = N_d(t)\mathbf{v}(t)$ derived via the shifted-Gaussian Schwinger-Dyson elimination for simple labelled $d$-regular graphs ($4 \le d \le 10$), together with the graph counts $r_d(n)$ reproduced by the resulting $O(n)$ recurrence. Here, $\mathcal{R}_d$ is the number of master correlators and $\deg N_d \equiv \max_{a,b} \deg (N_d)_{ab}$.}
\label{tab:gm-summary}
\end{table}
Based on these ideas, a small software package has been prepared by the publicly accessible version of Google Gemini AI, and then debugged
under the supervision of the human author. This implementation has been made available to the public \cite{repo}. The performance has been tested for $d=4,\dots,10$. (The case $d=3$ is reducible to a one-dimensional
integral using the techniques presented here, and is too simple to bother.) In all those cases where prior information is available \cite{OEIS,IM25},
the results are validated. A summary of the recurrence parameters is given in Table~\ref{tab:gm-summary}.
An important source on the state-of-the-art is the online repository associated with the recent preprint \cite{IM25},
which only includes graph counts up to $n=90$ for $d=9$ and up to $n=75$ for $d=10$.
Without any Herculean effort, and in a few hours of computation time on an ordinary desktop computer,
using the recursion designed here, one deduces the following two numbers beyond the previously existing computational limits:
at $d = 9$ and $n = 100$, we obtain the exact $570$-digit count
\begin{align*}
r_{9}(100) &= 156566685187171891526264469211081345642341584762769317010222399885171 \\
&\quad 991107010896137202445080174496501326499843850686240421255727984249423 \\
&\quad 226206624406333571710110783820428203765117773073476630787318175164513 \\
&\quad 075444429938697216077499532119906610266066651467440224540506259595944 \\
&\quad 976431316817504684012250619643249243262773089875657234441477860745650 \\
&\quad 579981762608985187116202657406293687073740315444603562781503170373016 \\
&\quad 360654233994071345871440189000857306932502511231449023291728586477457 \\
&\quad 625268368045744678724246451303358822122588191309711027528707027925688 \\
&\quad 323505109058499700,
\end{align*}
while at  $d = 10$ and $n=100$, we obtain the exact $616$-digit count
\begin{align*}
r_{10}(100) &= 801777308205742115189342893116274195778174508657758821518100733527509 \\
&\quad 159353027062443163683716689324281442871195349897279242756652390700533 \\
&\quad 297926492606445500254465443547324522399925916795682078630500876951088 \\
&\quad 365410972263738263288687102017763082137264577493325516251780941191479 \\
&\quad 461848874710821361751908028706429851287898296547800669418802150523932 \\
&\quad 621491243803408123004766955016015551467559236411620664053405412242723 \\
&\quad 573316510335087793851594105634479504541968594550242568646816050688489 \\
&\quad 934247479164798098728829896144803748962209193012037115489917560432838 \\
&\quad 8852925893471674694229212726830760364784790690980260272119951196,
\end{align*}
both of which\footnote{Following the initial release of this article as a preprint, the implementation of the modular arithmetic for running the iterations
has been substantially optimized by Google Gemini (without changing the derivation of the underlying differential system). As a result, one can comfortably
proceed up to $n=500$, obtaining $r_9(500)=124897...68750$ with 4455 digits and $r_{10}(500)=13714...88326$ with 4872 digits.} can be verified against the Isaev-McKay asymptotic formulas \cite{IM25}.


\section{Discussion}\label{sec:conc}

We have shown how the Schwinger-Dyson equations of zero-dimensional polynomial quantum field theories, paired with classical commutative Gr\"obner basis elimination, provide a ready-to-use and effective engine for deriving fixed-width linear recurrences that close on a finite set of correlators (moments). All other correlators
can be afterwards expressed, if necessary, through this master set. When applied to $d$-regular graph enumeration, relying on the Gaussian representation of \cite{EH,CdP23,dP24}, coupled with modular arithmetic and rational reconstruction, this approach pushes exact recurrence derivations to $d=9$ and $d=10$, beyond the reach of general creative telescoping methods \cite{CM26,BCL26} representing the current state-of-the-art. Then, from a basic desktop computer run over a handful of hours, one extracts the graph counts on 500 vertices mentioned in section~\ref{sec:results} that have been unreachable up until now.

The implications of the simple setup unveiled here deserve being explored further: In application to quantum field theory topics, the central question is whether
any of our insights, tied in the current presentation to finite-dimensional integrals, extend to full realistic quantum field theories defined by functional integrals.
While such aspirations may seem too far-fetched, it is reassuring that some similar holonomic-function techniques have emerged in the literature on 
specialized computations of higher-order Feynman diagrams. Could it be possible to trace some of these structures back to some holonomic properties of the original
functional integral from which the Feynman diagrams had been derived?

Closer to the setup of section~\ref{sec:SD-grobner}, one could consider, much in the spirit of \cite{MSh,Gurau}, integrals of the form
\beq
Z(\mathbf{T})=\int d\mathbf{x}\, e^{-\mathbf{x}^2+\mathbf{T}\mathbf{x}^4},
\eeq
where $\mathbf{T}$ is a tensor with components $T_{ijkl}$ and $\mathbf{T}\mathbf{x}^4=\sum_{ijkl}T_{ijkl}x_ix_jx_kx_l$. One could evidently consider higher-rank tensors instead, and other related formulations. The function $Z$ is then made of tensor invariants with respect to the rotation group, and it would be interesting to get better control over its properties. In \cite{MSh}, some similar integrals are recast in terms of hypergeometric functions, naturally connecting to the topic of having them satisfy linear differential equations. What are the implications, in that setting, of the ideas advocated here?

In enumerative combinatorics, the usefulness of elimination theory has long been known and the associated machinery is fully developed.
What seems less well-appreciated is that Gaussian integral representations (fundamentally, a reversal of pairing Feynman diagram drawings to Gaussian integral computations in conventional quantum field theory) can be very useful. Section~\ref{sec:SD-regular} provides a practical validation of this claim.
Gaussian integral representations will extend straightforwardly to counting multigraphs, directed graphs, and other generalizations,
also with an explicit specification of the total number of independent loops.
With extra effort, one may hope to adapt them to the problem of counting graphs with a prescribed number of triangles \cite{tri1}
and other motifs, which is also of considerable interest in statistical physics of networks \cite{tri2,tri3}. Other features like planarity or various degrees of connectedness may also be incorporated. Much remains to be explored.

\section*{Acknowledgments}

The author is grateful to Carmen Amarra, Misha Isaev, Manuel Loquias and \'Elie de Panafieu for stimulating discussions on graph combinatorics.

This work has essentially relied on the assistance of the publicly accessible Gemini AI agent (Google).
The theoretical stage was set by the human author based on his previous work on Gaussian representations for graph counting \cite{EH},
further stimulated by the applications of `creative telescoping' to this type of problems in \cite{CM26,BCL26}, and the recent article \cite{Schwarz}
on recursive generation of QFT Green's functions. The relevant connections to Schwinger-Dyson equations and the general idea of master monomial closure were also provided,
but the concrete Gr\"obner-basis strategy was suggested by the AI agent, who also went on to supply all the numerical programming and modular arithmetic implementation using SageMath and C++. The AI agent has also been very helpful with describing the Gr\"obner reduction in an elementary language adapted
for the theoretical physics context, and providing an initial draft of the paper, while showing great appreciation of the contributions of the human author and claiming
that, on its own and without the human inputs on the Gaussian representation of \cite{EH} and Schwinger-Dyson relations, it would have merely ``diagnosed why $r_{10}(100)$ was beyond the known literature.'' All aspects of the logic and analytics have been critically reviewed by the human author,
also in conversations with the AI  agent, while the numerical implementation has been tested and debugged thoroughly by the AI agent under human supervision.
The human author is thereby convinced that the presented framework is sound, while the ultimate test comes from being able to proceed with beyond-the-state-of-the-art ultrasensitive computations (recovering multidigit integers from rational recursion), where a microscopic mistake would have disrupted the performance completely, and where multiple intermediate checks are available from smaller-size computations in the past literature.


\appendix

\section{Gr\"obner-free reduction for a two-variable even quartic}\label{app:quartic}

To illustrate the box-basis reduction of section~\ref{subsec:generic-box} without Gr\"obner bases, consider the two-variable Gaussian-anchored integral
\begin{equation}
Z(t) = \int_{-\infty}^{\infty} \frac{dx\,dy}{2\pi}\, e^{S(t; x, y)},
\end{equation}
with a general $\mathbb{Z}_2 \times \mathbb{Z}_2$-symmetric quartic interaction ($M = 4$, $k = 2$),
\begin{equation}
S(t; x, y) = -\frac{x^2 + y^2}{2} + t\,V(x, y), \qquad V(x, y) = A x^4 + B x^2 y^2 + C y^4.
\end{equation}
The classical equations of motion $E_x = \partial_x S = 0$ and $E_y = \partial_y S = 0$ read
\begin{equation}
E_x = -x + t(4A x^3 + 2B x y^2), \qquad E_y = -y + t(2B x^2 y + 4C y^3).
\end{equation}
Their top-degree parts have $(x, y) = (0, 0)$ as their unique common complex root whenever $A \neq 0$, $C \neq 0$, and
\begin{equation}
\delta \equiv 4AC - B^2 \neq 0.
\end{equation}
Under this nondegeneracy condition, the master monomials restrict from the $3 \times 3$ box $\{x^{p_1} y^{p_2} : 0 \le p_1, p_2 \le 2\}$ to its four $\mathbb{Z}_2 \times \mathbb{Z}_2$-invariant elements:
\begin{equation}
\mathcal{B} = \{1,\; x^2,\; y^2,\; x^2 y^2\}, \qquad
\mathbf{v}(t) = \begin{pmatrix} v_0(t) \\ v_1(t) \\ v_2(t) \\ v_3(t) \end{pmatrix}
\equiv \begin{pmatrix} \langle 1 \rangle_t \\ \langle x^2 \rangle_t \\ \langle y^2 \rangle_t \\ \langle x^2 y^2 \rangle_t \end{pmatrix}.
\end{equation}

\subsection*{Degree-by-degree Schwinger-Dyson elimination}

Applying the \SD{} identities $\langle (\partial_x + E_x)f \rangle_t = 0$ and $\langle (\partial_y + E_y)f \rangle_t = 0$ at total degrees $w = 4, 6, 8$ reduces all higher even-even correlators onto $\mathcal{B}$ by elementary linear algebra:

\paragraph{Degree $4$ ($2 \times 2$ system).}
The two identities $\langle \mathcal{D}_x(x) \rangle_t = 0$ and $\langle \mathcal{D}_y(y) \rangle_t = 0$ decouple directly into
\begin{equation}
\langle x^4 \rangle_t = \frac{v_1 - v_0}{4At} - \frac{B}{2A}\,v_3, \qquad
\langle y^4 \rangle_t = \frac{v_2 - v_0}{4Ct} - \frac{B}{2C}\,v_3.
\end{equation}

\paragraph{Degree $6$ ($4 \times 4$ system).}
Solving the four identities $\langle \mathcal{D}_x(x^3) \rangle_t = \langle \mathcal{D}_x(x y^2) \rangle_t = \langle \mathcal{D}_y(x^2 y) \rangle_t = \langle \mathcal{D}_y(y^3) \rangle_t = 0$ and substituting the degree-$4$ reductions above yields
\begin{align}
\langle x^6 \rangle_t &= -\frac{v_0}{16 A^2 t^2} + \frac{\delta + 4A(3AC - 2B^2)t}{16 A^2 \delta\, t^2}\,v_1 - \frac{BC}{2A\delta\, t}\,v_2 + \frac{B(2\delta + 2AB - B^2)}{8 A^2 \delta\, t}\,v_3, \nn\\
\langle x^4 y^2 \rangle_t &= -\frac{B}{2\delta\, t}\,v_1 + \frac{C}{\delta\, t}\,v_2 + \frac{B - 2C}{2\delta\, t}\,v_3, \nn\\
\langle x^2 y^4 \rangle_t &= \frac{A}{\delta\, t}\,v_1 - \frac{B}{2\delta\, t}\,v_2 + \frac{B - 2A}{2\delta\, t}\,v_3, \nn\\
\langle y^6 \rangle_t &= -\frac{v_0}{16 C^2 t^2} - \frac{AB}{2C\delta\, t}\,v_1 + \frac{\delta + 4C(3AC - 2B^2)t}{16 C^2 \delta\, t^2}\,v_2 + \frac{B(2\delta + 2BC - B^2)}{8 C^2 \delta\, t}\,v_3.
\end{align}

\paragraph{Degree $8$ ($6 \times 5$ system of rank $5$).}
Solving the six identities $\langle \mathcal{D}_x(x^5) \rangle_t = \langle \mathcal{D}_x(x^3 y^2) \rangle_t = \langle \mathcal{D}_x(x y^4) \rangle_t = \langle \mathcal{D}_y(x^4 y) \rangle_t = \langle \mathcal{D}_y(x^2 y^3) \rangle_t = \langle \mathcal{D}_y(y^5) \rangle_t = 0$ expresses $\langle x^8 \rangle_t$, $\langle x^6 y^2 \rangle_t$, $\langle x^4 y^4 \rangle_t$, $\langle x^2 y^6 \rangle_t$, and $\langle y^8 \rangle_t$ in terms of $\mathbf{v}(t)$ in the same manner.

\subsection*{First-order $4 \times 4$ ODE system and one-step recurrence}

Substituting these reductions into $\mathbf{v}'(t) = \langle V(x,y)\,\mathbf{b}(x,y) \rangle_t$ and clearing the common denominator $16 A C \delta\, t^2$ gives the exact first-order system
\begin{equation}\label{eq:quartic-ode}
16 A C \delta\, t^2\,\frac{d\mathbf{v}(t)}{dt} = \big(N_0 + t\,N_1\big)\,\mathbf{v}(t),
\end{equation}
where
\begin{equation}
N_0 = \begin{pmatrix}
0 & 0 & 0 & 0 \\
-C\delta & C\delta & 0 & 0 \\
-A\delta & 0 & A\delta & 0 \\
0 & 2AC(B - 2A) & 2AC(B - 2C) & -4AC(B - A - C)
\end{pmatrix},
\end{equation}
and
\begin{equation}
N_1 = \begin{pmatrix}
-8AC\delta & 4AC\delta & 4AC\delta & 0 \\
0 & -16AC\delta & 0 & 2C(2A - B)\delta \\
0 & 0 & -16AC\delta & 2A(2C - B)\delta \\
0 & 0 & 0 & -24AC\delta
\end{pmatrix}.
\end{equation}

Expanding $\mathbf{v}(t) = \sum_{n=0}^{\infty} \mathbf{v}_n\, t^n / n!$ with the free Gaussian initial vector $\mathbf{v}_0 = (1, 1, 1, 1)^T$, matching coefficients of $t^{n+1}/(n+1)!$ in the first row and $t^n/n!$ in rows $2$--$4$ of \eqref{eq:quartic-ode} yields a closed $3 \times 3$ linear system determining $(v_{1,n}, v_{2,n}, v_{3,n})$ directly from $\mathbf{v}_{n-1}$ at every $n \ge 1$ (since $N_0$ vanishes on the first row and has invertible lower-right $3 \times 3$ block whenever $A, C, \delta \neq 0$), followed immediately by the first-row update
\begin{equation}
(4n + 2)\,v_{0,n} = v_{1,n} + v_{2,n}.
\end{equation}
In particular, at $n = 1$ this reproduces $v_{0,1} = 3A + B + 3C$, $v_{1,1} = 15A + 3B + 3C$, $v_{2,1} = 3A + 3B + 15C$, and $v_{3,1} = 15A + 9B + 15C$, generating the perturbative expansion of $Z(t)$ recursively in $O(n)$ algebraic operations.


\section{Explicit analysis for the case $d=4$}\label{app:d4}

For regular graphs with $d=4$, the processing of the \SD{} equations is compact enough to present it analytically, as we proceed to demonstrate:

\subsection*{Gaussian elimination of linear variables $x_3$ and $x_4$}

For $d = 4$, the cycle-index polynomial $P_4(x_1, x_2, x_3, x_4) = [u^4]\exp\bigl(\sum_{m=1}^4 x_m u^m/m\bigr)$ is
\begin{equation}
P_4(x_1, x_2, x_3, x_4) = \frac{x_1^4}{24} + \frac{x_1^2 x_2}{4} + \frac{x_2^2}{8} + \frac{x_1 x_3}{3} + \frac{x_4}{4}.
\end{equation}
With the weights $w(x_m) = m$, every term in $P_4$ has total weight $4$. The 4-regular graph counting function
\begin{equation}
R_4(t) = \int d\mathbf{x}\,\exp\!\left( -\frac{x_1^2}{2} - \frac{(x_2 - i)^2}{4} - \frac{x_3^2}{6} - \frac{(x_4 + i)^2}{8} + t\,P_4(x_1, i x_2, x_3, i x_4) \right),
\end{equation}
where $d\mathbf{x}=dx_1\,dx_2\,dx_3\,dx_4/(2\pi)^2\sqrt{4!}$.
Because $x_3$ and $x_4$ ($m > d/2 = 2$) appear strictly linearly in the interaction, completing the square integrates both variables out, leaving behind
\beq
R_4(t) = \int \frac{dx_1\,dx_2}{2\pi\sqrt{2}}\,e^{-S_{\mathrm{eff}}(x_1, x_2; t)},
\eeq 
with effective action
\begin{equation}
S_{\mathrm{eff}}(x_1, x_2; t) = \frac{x_1^2}{2} + \frac{(x_2 - i)^2}{4} - t\left(\frac{x_1^4}{24} - \frac{i x_1^2 x_2}{4} - \frac{x_2^2}{8} + \frac{1}{4}\right) - \frac{t^2}{2}\left(\frac{x_1^2}{3} - \frac{1}{4}\right).
\end{equation}
Differentiating $R_4(t)$ with respect to $t$ gives $R_4'(t) = \langle \Sigma_{\mathrm{eff}}(x_1, x_2; t) \rangle_t$, where
\begin{equation}
\Sigma_{\mathrm{eff}}(x_1, x_2; t) = -\frac{\partial S_{\mathrm{eff}}}{\partial t} = \frac{x_1^4}{24} - \frac{i x_1^2 x_2}{4} - \frac{x_2^2}{8} + \frac{1}{4} + t\left(\frac{x_1^2}{3} - \frac{1}{4}\right).
\end{equation}

\subsection*{Schwinger-Dyson equations}

Because $e^{-S_{\mathrm{eff}}(x_1, x_2; t)}$ decays rapidly at infinity when $t=0$, integrating the total derivatives $\partial(f\,e^{-S_{\mathrm{eff}}})/\del{x_1}$ and $\partial(f\,e^{-S_{\mathrm{eff}}})/\del{x_2}$ for any polynomial $f(x_1, x_2)$ gives the Schwinger-Dyson identities for formal power series in $t$: $\langle \mathcal{D}_1 f \rangle_t = 0$ and $\langle \mathcal{D}_2 f \rangle_t = 0$, with differential operators
\begin{align}
\mathcal{D}_1 &= \frac{\partial S_{\mathrm{eff}}}{\partial x_1} - \frac{\partial}{\partial x_1} = x_1 - t\left(\frac{x_1^3}{6} - \frac{i x_1 x_2}{2}\right) - \frac{t^2 x_1}{3} - \frac{\partial}{\partial x_1}, \\
\mathcal{D}_2 &= \frac{\partial S_{\mathrm{eff}}}{\partial x_2} - \frac{\partial}{\partial x_2} = \frac{x_2 - i}{2} + \frac{t(i x_1^2 + x_2)}{4} - \frac{\partial}{\partial x_2}.
\end{align}
We order monomials $x_1^{\alpha_1} x_2^{\alpha_2}$ by weighted degree $W = \alpha_1 + 2\alpha_2$ with reverse lexicographic tie-breaking ($x_2 > x_1$). In $\mathcal{D}_1$ and $\mathcal{D}_2$, the derivative-free terms have top weight $3$ and $2$, respectively, whereas $\partial/\del{x_1}$ and $\partial/\del{x_2}$ lower weight by $1$ and $2$.

\subsection*{Gr\"obner basis and master monomials}

Dropping the derivative terms from the \SD{} operators leaves
\begin{equation}
E_1 = -\frac{t x_1^3}{6} + \frac{i t x_1 x_2}{2} + \frac{(3 - t^2) x_1}{3}, \qquad
E_2 = \frac{(t + 2) x_2}{4} + \frac{i t x_1^2}{4} - \frac{i}{2}.
\end{equation}
For generic $t \neq 0, 1, -2$, normalizing $E_2$ and eliminating $x_1 x_2$ from $E_1$ produces the Gr\"obner basis $\{g_1, g_2\}$ under the weight-graded reverse lexicographic order:
\begin{equation}
g_1(x_1, x_2) = x_2 + \frac{i(t x_1^2 - 2)}{t + 2}, \qquad
g_2(x_1, x_2) = x_1^3 + \frac{(t - 1)(t + 2) x_1}{t}.
\end{equation}
Their leading monomials are $\{x_2, x_1^3\}$. A monomial $x_1^{p_1} x_2^{p_2}$ is divisible by neither $x_2$ nor $x_1^3$ if and only if $p_2 = 0$ and $p_1 \in \{0, 1, 2\}$, leaving $\{1, x_1, x_1^2\}$. Because $S_{\mathrm{eff}}(-x_1, x_2; t) = S_{\mathrm{eff}}(x_1, x_2; t)$, all odd-weight expectations vanish identically ($\langle x_1 \rangle_t = 0$), leaving the two basic monomials ($\mathcal{R}_4 = 2$):
\begin{equation}
\mathcal{B} = \{e_0, e_1\} = \{1,\, x_1^2\}.
\end{equation}

\subsection*{Explicit reduction to the $2 \times 2$ linear ODE system}

We then explore the $t$-dependence of the master correlator vector $\mathbf{v}(t) = (v_0(t), v_1(t))^T$ with $v_0(t) = \langle 1 \rangle_t = R_4(t)$ and $v_1(t) = \langle x_1^2 \rangle_t$. Differentiating with respect to $t$ gives $v_0'(t) = \langle \Sigma_{\mathrm{eff}} \rangle_t$ and $v_1'(t) = \langle \Sigma_{\mathrm{eff}} x_1^2 \rangle_t$. Solving the weight-$2$ and weight-$4$ Schwinger-Dyson identities $\langle \mathcal{D}_2(1) \rangle_t = 0$, $\langle \mathcal{D}_1(x_1) \rangle_t = 0$, $\langle \mathcal{D}_2(x_1^2) \rangle_t = 0$, and $\langle \mathcal{D}_2(x_2) \rangle_t = 0$ expresses all non-basis correlators up to weight~$4$ linearly in terms of $\langle 1 \rangle_t$ and $\langle x_1^2 \rangle_t$:
\begin{align}
\langle x_2 \rangle_t &= \frac{i\big(t\,\langle x_1^2 \rangle_t + 2\langle 1 \rangle_t\big)}{t + 2}, \nn\\
\langle x_1^4 \rangle_t &= \frac{(t^3 + 2t^2 - 6t - 6)\,\langle x_1^2 \rangle_t + 3(t + 2)\langle 1 \rangle_t}{t(t - 1)},\nn \\
\langle x_1^2 x_2 \rangle_t &= \frac{i\big((t - 2)(t + 2)\,\langle x_1^2 \rangle_t + 3\langle 1 \rangle_t\big)}{t - 1}, \nn\\
\langle x_2^2 \rangle_t &= -\frac{t(t^3 + 2t^2 - 2t - 10)\,\langle x_1^2 \rangle_t - (t^2 - 6t - 4)\langle 1 \rangle_t}{(t - 1)(t + 2)^2}.\nn
\end{align}
Similarly, solving the weight-$6$ identities $\langle \mathcal{D}_1(x_1^3) \rangle_t = \langle \mathcal{D}_1(x_1 x_2) \rangle_t = \langle \mathcal{D}_2(x_1^4) \rangle_t = \langle \mathcal{D}_2(x_1^2 x_2) \rangle_t = \langle \mathcal{D}_2(x_2^2) \rangle_t = 0$ expresses all weight-$6$ correlators linearly in terms of $\langle 1 \rangle_t$ and $\langle x_1^2 \rangle_t$:
\begin{align}
\langle x_1^6 \rangle_t &= \frac{(t^6 + 4t^5 - 8t^4 - 27t^3 + 21t^2 + 54t + 36)\,\langle x_1^2 \rangle_t + 3(t + 2)(t^3 + 2t^2 - 6t - 6)\langle 1 \rangle_t}{t^2(t - 1)^2}, \nn\\
\langle x_1^4 x_2 \rangle_t &= \frac{i\big[(t^5 + 2t^4 - 10t^3 - 5t^2 + 15t + 24)\,\langle x_1^2 \rangle_t + 3(t - 2)(t + 2)^2\langle 1 \rangle_t\big]}{t(t - 1)^2}, \nn\\
\langle x_1^2 x_2^2 \rangle_t &= -\frac{(t^5 + 2t^4 - 8t^3 - 11t^2 + 15t + 28)\,\langle x_1^2 \rangle_t + 3(t^3 + 2t^2 - 2t - 10)\langle 1 \rangle_t}{(t - 1)^2(t + 2)}, \nn\\
\langle x_2^3 \rangle_t &= -\frac{i\,t(t^6 + 4t^5 - 2t^4 - 33t^3 - 15t^2 + 66t + 60)\,\langle x_1^2 \rangle_t}{(t - 1)^2(t + 2)^3}\nn\\
&\hspace{5.5cm}- \frac{i(3t^5 + 12t^4 - 12t^3 - 28t^2 - 16t - 40)\langle 1 \rangle_t}{(t - 1)^2(t + 2)^3}.\nn
\end{align}
Substituting these reductions into $v_0'(t) = \langle \Sigma_{\mathrm{eff}} \rangle_t$ and $v_1'(t) = \langle \Sigma_{\mathrm{eff}} x_1^2 \rangle_t$ and clearing the common denominator $4t^2(t - 1)^2(t + 2)^2$ yields the exact $2 \times 2$ linear ODE system
\begin{equation}
4t^2(t - 1)^2(t + 2)^2\,\frac{d}{dt}\begin{pmatrix} v_0(t) \\ v_1(t) \end{pmatrix} = \begin{pmatrix} N_{00}(t) & N_{01}(t) \\ N_{10}(t) & N_{11}(t) \end{pmatrix}\begin{pmatrix} v_0(t) \\ v_1(t) \end{pmatrix},
\end{equation}
where
\begin{align}
N_{00}(t) &= -t^7 - t^6 + 2t^5 - 2t^4 - 6t^3 + 12t^2 - 4t, \nn\\
N_{01}(t) &= t^7 + t^6 - 2t^5 + 2t^4 + 6t^3 - 12t^2 + 4t,\nn \\
N_{10}(t) &= 3t^6 + 12t^5 + 12t^4 + 6t^3 + 36t^2 + 36t - 24, \nn\\
N_{11}(t) &= t^8 + 3t^7 - 3t^6 - 16t^5 - 12t^4 - 6t^3 - 36t^2 - 36t + 24.\nn
\end{align}

\subsection*{Initial conditions and the resolved coupled recurrence}

At $t = 0$, the free Gaussian expectation values give $\mathbf{v}(0) = (v_0(0), v_1(0))^T = (1, 1)^T$. Expanding
\begin{equation}
v_0(t) = \sum_{n=0}^{\infty} r_n\,\frac{t^n}{n!}, \qquad v_1(t) = \sum_{n=0}^{\infty} u_n\,\frac{t^n}{n!},
\end{equation}
and equating coefficients of $\frac{t^{n+1}}{(n+1)!}$ in the first row and $\frac{t^n}{n!}$ in the second row of $D(t)\,\mathbf{v}'(t) = N(t)\,\mathbf{v}(t)$, using the falling factorial notation $n^{\underline{k}} = n(n - 1)\cdots(n - k + 1)$ with $n^{\underline{0}} = 1$, we solve the resulting $2 \times 2$ linear system for $(r_n, u_n)$ to obtain
\begin{align}
r_n &= \frac{1}{96n}\Bigl[
(96n - 60)\,n^{\underline{1}}\,r_{n-1}
+ (72n - 216)\,n^{\underline{2}}\,r_{n-2}
+ (-48n + 126)\,n^{\underline{3}}\,r_{n-3} \nonumber \\
&\quad + (-24n + 96)\,n^{\underline{4}}\,r_{n-4}
- 18\,n^{\underline{5}}\,r_{n-5}
- 9\,n^{\underline{6}}\,r_{n-6}
+ (16n - 52)\,n^{\underline{1}}\,u_{n-1} \nonumber \\
&\quad + (-16n + 104)\,n^{\underline{2}}\,u_{n-2}
+ (-12n + 54)\,n^{\underline{3}}\,u_{n-3}
+ (8n - 32)\,n^{\underline{4}}\,u_{n-4} \nonumber \\
&\quad + (4n + 2)\,n^{\underline{5}}\,u_{n-5}
+ 9\,n^{\underline{6}}\,u_{n-6}
- 3\,n^{\underline{7}}\,u_{n-7}
- n^{\underline{8}}\,u_{n-8}
\Bigr],
\end{align}
and
\begin{align}
u_n &= \frac{1}{96n}\Bigl[
(-48n - 60)\,n^{\underline{1}}\,r_{n-1}
+ (-72n - 216)\,n^{\underline{2}}\,r_{n-2}
+ (-72n + 126)\,n^{\underline{3}}\,r_{n-3} \nonumber \\
&\quad + (-72n + 96)\,n^{\underline{4}}\,r_{n-4}
+ (-48n - 18)\,n^{\underline{5}}\,r_{n-5}
+ (-12n - 9)\,n^{\underline{6}}\,r_{n-6} \nonumber \\
&\quad + (64n^2 + 96n - 52)\,n^{\underline{1}}\,u_{n-1}
+ (-64n^2 + 256n + 104)\,n^{\underline{2}}\,u_{n-2} \nonumber \\
&\quad + (-48n^2 + 156n + 54)\,n^{\underline{3}}\,u_{n-3}
+ (32n^2 - 72n - 32)\,n^{\underline{4}}\,u_{n-4} \nonumber \\
&\quad + (16n^2 - 12n + 2)\,n^{\underline{5}}\,u_{n-5}
+ (12n + 9)\,n^{\underline{6}}\,u_{n-6} \nonumber \\
&\quad + (-12n - 3)\,n^{\underline{7}}\,u_{n-7}
+ (-4n - 1)\,n^{\underline{8}}\,u_{n-8}
\Bigr].
\end{align}
Note that no extra initial conditions are needed to launch this recursion apart from $r_0 = u_0 = 1$,
as all terms at negative indices are cancelled automatically by the vanishing of the falling factorials $n^{\underline{k}}$ for $n < k$. Evaluating the recurrence gives $r_1=r_2 = r_3 = r_4 = 0$ 
before correctly proceeding with $r_5 = 1$, $r_6 = 15$, $r_7 = 465$, $r_8 = 19355$, $r_9 = 1024380$, $r_{10} = 66462606$, ... \cite{OEIS}.


\section{Modular arithmetic and rational reconstruction}\label{app:modular}

When deriving the polynomials $D_d(t)$ and $N_d(t)$ for $d\ge 6$, naive symbolic Gaussian elimination over rational functions $\mathbb{Q}(t)$ slows down dramatically because intermediate fractions accumulate huge numerators and denominators that cancel out only at the very end. Modular arithmetic proves useful in avoiding this bottleneck:

\subsection*{Working modulo a prime number}

Instead of computing with rational numbers $a/b \in \mathbb{Q}$ (whose numerators and denominators can grow to hundreds of digits when Gaussian elimination is used to solve the systems of linear equations occurring at the intermediate stages), choose a large prime number $q$ that fits inside a standard $64$-bit computer word, such as $q = 2^{61}-1$. Modulo $q$, every integer is replaced by its remainder in $\{0, 1, \dots, q-1\}$, and division by any nonzero number $b$ is simply multiplication by its modular inverse $b^{-1} \pmod q$. Because every number stays strictly smaller than $q < 2^{64}$, addition, subtraction, multiplication, and division take a single hardware instruction and never suffer from memory or digit-length growth.

Similarly, instead of carrying a symbolic variable $t$ through Gaussian elimination, we substitute concrete integer values $t = t_1, t_2, \dots, t_K$ one at a time. At each fixed value $t=t_i$ and prime $q_\ell$, the Schwinger-Dyson linear system becomes a sparse matrix of $64$-bit integers modulo $q_\ell$, which can be solved in parallel across CPU cores to obtain the entries $M_{ab}(t_i) = [D_d(t_i)^{-1} N_d(t_i)]_{ab}  \bmod q_\ell$.

\subsection*{Reconstructing exact rational numbers via the Chinese Remainder Theorem}

Suppose we repeat this calculation for several distinct $61$-bit primes $q_1, q_2, \dots, q_L$ (for instance, $L=8$ primes give a combined modulus $M = q_1 q_2 \cdots q_L > 2^{480}$, more than $140$ decimal digits). Given the remainders $c_\ell \equiv c \pmod{q_\ell}$ for $\ell=1,\dots,L$, the classic \emph{Chinese Remainder Theorem} combines them into a single integer $C \in \{0, 1, \dots, M-1\}$ that leaves the exact same remainder modulo every $q_\ell$.

How do we recover a fraction $a/b \in \mathbb{Q}$ from the single integer $C \equiv a\,b^{-1} \pmod M$? Provided $2|a|b < M$ (for example, whenever $|a|$ and $b$ are both smaller than $\sqrt{M/2}$, which is guaranteed once $L$ is large enough), there is a \emph{unique} fraction $a/b$ in lowest terms satisfying $a \equiv C\,b \pmod M$. Running Euclid's greatest-common-divisor algorithm on $(M, C)$ and stopping as soon as the remainder drops below $\sqrt{M/2}$ produces $(a, b)$ immediately. This step, known as \emph{rational reconstruction}, lifts every modular evaluation $M_{ab}(t_i) \bmod q_\ell$ back to the exact rational number $M_{ab}(t_i) \in \mathbb{Q}$.

\subsection*{Reconstructing rational functions of $t$}

Once we have the exact rational values $y_i = M_{ab}(t_i) \in \mathbb{Q}$ at $K$ distinct points $t_1,\dots,t_K$, we wish to recover the rational function $P(t)/Q(t) \in \mathbb{Q}(t)$ (where $\deg P \le D_{\mathrm{num}}$, $\deg Q \le D_{\mathrm{den}}$, and $D_{\mathrm{num}} + D_{\mathrm{den}} < K$). This is the exact polynomial analogue of rational reconstruction:
\begin{enumerate}
\item Construct the unique Lagrange interpolating polynomial $L(t) \in \mathbb{Q}[t]$ of degree $< K$ satisfying $L(t_i) = y_i$ for all $i=1,\dots,K$, together with the vanishing polynomial $W(t) = \prod_{i=1}^{K}(t - t_i)$.
\item Because $P(t)/Q(t)$ agrees with $L(t)$ at every root of $W(t)$, we have $P(t) \equiv L(t)\,Q(t) \pmod{W(t)}$. Running the polynomial Euclidean algorithm on $(W(t), L(t))$ and stopping when the remainder polynomial reaches degree $\le D_{\mathrm{num}}$ yields the exact numerator $P(t)$ and denominator $Q(t)$ uniquely.
\end{enumerate}
Clearing the least common multiple $D_d(t)$ of all denominator polynomials $Q(t)$ across the $\mathcal{R}_d \times \mathcal{R}_d$ matrix produces the exact integer-coefficient polynomials $D_d(t)$ and $N_d(t)$.

\subsection*{Multi-prime modular evaluation of the recurrence}

The exact same modular strategy accelerates the evaluation of the vector recurrence \eqref{eq:vec-rec} up to large $n$. Using the Bender-Canfield asymptotic growth formula \cite{BC} 
$$
r_d(n) \approx \frac{\sqrt{2}\,e^{-(d^2-1)/4}\,(dn)^{dn/2}}{e^{dn/2}\,(d!)^n},
$$ 
we know ahead of time how many bits $r_d(n)$ can have at any $n\le N$ (leading growth $\frac{dn}{2} \log_2 n$ bits). Choosing enough $61$-bit primes $q_1,\dots,q_L$ so that $M = \prod_{\ell=1}^L q_\ell > 2\, r_d(N)$, we solve the banded linear recurrence system \eqref{eq:vec-rec} modulo each prime $q_\ell$ in parallel, initializing $\mathbf{v}_0,\dots,\mathbf{v}_s$ directly from the low-order Gaussian integrals of \eqref{eq:Rd-eff} to bypass the $t=0$ degeneracy of $D_d(t)$. We then verify consistency against an independent check prime, and reconstruct the exact multiprecision integers $r_d(n)$ via the Chinese Remainder Theorem.


\end{document}